\documentstyle[seceq,preprint]{jpsj}

\newcommand{\nolig}{\textcompwordmark} \newcommand{\eqref}[1]{(\ref{#1})}
\newcommand{\Ge}{CeCu$_2$Ge$_2$ } \newcommand{\Si}{CeCu$_2$Si$_2$ }
\newcommand{\ep}{\epsilon} \newcommand{\Ufc}{U_{\rf\rc}}
\newcommand{\sqNL}{\sqrt{\NL}}
\newcommand{\sH}{{\mathcal{H}}} \newcommand{\sD}{{\mathcal{D}}}
\newcommand{\rc}{{\rm c}} \newcommand{\rf}{{\rm f}}
\newcommand{\rd}{{\rm d}} \newcommand{\NL}{N_{\rm L}}
\newcommand{\ex}{{\rm e}} \newcommand{\mi}{{\rm i}}
\newcommand{\rk}{{\rm{k}}} \newcommand{\rmq}{{\rm{q}}}
\newcommand{\bk}{{\mib{k}}} \newcommand{\bks}{{\mibs{k}}}
\newcommand{\bq}{{\mib{q}}} \newcommand{\bqs}{{\mibs{q}}}
\newcommand{\bzero}{{\mib{0}}}
\newcommand{\phif}{\varphi^\rf} \newcommand{\phic}{\varphi^\rc}
\def\Tr{\mathop{{\rm{Tr}}}\nolimits}
\newcommand{\rhobar}{\bar{\rho}} \newcommand{\lambar}{\bar{\lambda}}
\newcommand{\phifbar}{\bar{\varphi}^\rf} 
\newcommand{\phicbar}{\bar{\varphi}^\rc}
\newcommand{\epbar}{\bar{\epsilon}} \newcommand{\Vbar}{\bar{V}}
\newcommand{\nbar}{\bar{n}}
\newcommand{\lamtil}{\tilde{\lambda}} \newcommand{\rhotil}{\tilde{\rho}}
\newcommand{\phiftil}{\tilde{\varphi}^\rf}
\newcommand{\phictil}{\tilde{\varphi}^\rc}
\newcommand{\rr}{{\rho\rho}} \newcommand{\rl}{{\rho\lambda}}
\newcommand{\rpc}{{\rho\phic}} \newcommand{\rpf}{{\rho\phif}}
\newcommand{\lr}{{\lambda\rho}} \newcommand{\lala}{{\lambda\lambda}}
\newcommand{\lpc}{{\lambda\phic}} \newcommand{\lpf}{{\lambda\phif}}
\newcommand{\pcr}{{\phic\rho}} \newcommand{\pcl}{{\phic\lambda}}
\newcommand{\pcpc}{{\phic\phic}} \newcommand{\pcpf}{{\phic\phif}}
\newcommand{\pfr}{{\phif\rho}} \newcommand{\pfl}{{\phif\lambda}}
\newcommand{\pfpc}{{\phif\phic}} \newcommand{\pfpf}{{\phif\phif}}
\newcommand{\Pic}{\Pi^{\rc}} \newcommand{\Pif}{\Pi^{\rf}}
\newcommand{\Picf}{\Pi^{\rc\rf}} \newcommand{\Pio}{\Pi_1}
\newcommand{\Pitw}{\Pi_2} \newcommand{\Pith}{\Pi_3}
\newcommand{\Gcc}{G_0^{\rc\rc}} \newcommand{\Gcf}{G_0^{\rc\rf}}
\newcommand{\Gfc}{G_0^{\rf\rc}} \newcommand{\Gff}{G_0^{\rf\nolig\rf}}
\def\runtitle{Enhanced Valence Fluctuations 
Caused by f-c Coulomb Interaction}
\def\runauthor{Yoshifumi {\sc Onishi} and Kazumasa {\sc Miyake}}

\title
{
Enhanced Valence Fluctuations Caused by f-c Coulomb 
Interaction in Ce-Based Heavy Electrons: \\
Possible Origin of Pressure-Induced Enhancement of 
Superconducting Transition Temperature in \Ge and Related Compounds
}

\author
{ 
Yoshifumi {\sc Onishi}\footnote{Present address: 
NEC Corporation, 4-1-1 Miyazaki, Miyamae-ku, Kawasaki, Kanagawa 216-8555, 
Japan.  
} 
and Kazumasa {\sc Miyake}
}

\inst
{
Division of Materials Physics, 
Department of Physical Science, Graduate School of Engineering Science \\
Osaka University, Toyonaka, Osaka 560-8531, Japan
}

\recdate
{
\today
}

\abst
{Properties of an extended periodic Anderson model with the f-c 
Coulomb interaction $\Ufc$ is studied as a model for \Ge and 
related compounds which is cosidered to exhibit a sharp valence change 
under pressure.  The problem is treated by extending the {\it slave-boson} 
and large-$N$ expansion method to treat the present system.  
It is shown that, as the f-level $\epsilon_{\rm f}$ is increased relative 
to the Fermi level, 
the sharp valence change is caused by the effect of $\Ufc$ with 
moderate strength of the order of the bandwith of conduction electrons.  
The superconducting transition temperature $T_\rc$ due to the 
valence-fluctuation exchange is estimated on the slave-boson fluctuation 
approximation.  In the model with spherical Fermi surface, 
$T_\rc$ exhibits sharp peak as a function of $\epsilon_{\rm f}$, simulating 
the effect of pressure, for the $d$-wave pairing channel.  
}

\kword
{
\Ge, f-c Coulomb interaction, valence transition, 
unconventional superconductivity
}

\begin{document}
\sloppy
\maketitle
\section{Introduction}\label{sec:1}
\Si is well known not only as the firstly discovered unconventional 
superconductor with 
$T_\rc \sim 0.7$ K at ambient pressure~\cite{rf:Steglich}, but as a 
remarkable pressure dependence of $T_\rc$ exhibiting a pronounced 
peak at $P\simeq 17$ GPa~\cite{rf:si1}.  The isostructural compound, \Ge, has a similar phase 
diagram in $P$-$T$ plane in which one can see the superconductivity 
appears after the magnetism is suppressed by pressure at 
$P\sim8$ GPa.~\cite{rf:Jac1,rf:Jac2,rf:kobayashi} 
These two compounds have similar physical properties, if the scale of the 
pressure for \Si is shifted by 7.6 GPa~\cite{rf:kita}.  
The mechanism of such prominent enhancement of $T_{\rm c}$ has not been 
understood so far.  A purpose of this paper is to discuss it 
from the view point that such phenomena is based on the sharp valence 
change of Ce ion.

Apart from the pronounced peak of $T_{\rm c}$, remarkable properties under 
pressure of these compounds are as follows:~\cite{rf:Jac1} 
1) The residual resistivity $\rho_{0}$ also exhibits a peak at around 
the pressure where $T_{\rm c}$ takes the maximum.  
2) The coefficient $A$ of the $T^2$-term of resistivity rapidly 
decreases by about two orders of magnitude at around the same 
pressure as $T_{\rm c}$ and $\rho_{0}$ exhibit peaks.
These phenomena may be attributed to a rapid change of the valence of 
Ce ion,~\cite{rf:MNO} 
while the origin of the superconductivity around magneic 
quantum critical point (QCP) is considered to be induced by the 
enhanced spin fluctuations.~\cite{rf:MSV}  
It is not only becasue the position of the sharp peak of $T_\rc$ is 
located 
far away from the pressure corresponding to QCP, but the valence of 
Ce ion seems to decrease rapidly there.  The latter fact is supported 
by the rapid decrease of the coefficient $A$ by about 10$^{-3}$ times, 
which implies that the Sommerfeld constant $\gamma$ decreases 
by more than 10$^{-1}$ times suggesting that the system is changed 
from the Kondo regime to the valence fluctuation (VF) regime rapidly.  
This is also supplemented by the rapid decrease of the so-called 
Kadowaki-Woods ratio $A/\gamma^{2}$ by 25 times there from the value 
of heavy electrons to that of the conventional transition 
metals.\cite{rf:Jac2,rf:KW,rf:Rice,rf:MMV}  

Recently, we have proposed that the physics behind these intriguing 
behaviors should be the enhanced valence fluctuations of Ce ion, and 
given a consistent explanation on the basis of a phenomenological model 
for the valence fluctuations.~\cite{rf:MNO} 
In the present paper we develope its microscopic theory on the basis 
of an extended periodic Anderson model with repulsion $\Ufc$ 
between f- and conduction electrons.  Without $\Ufc$, the valence change does 
not occur so drastically.  On the other hand, it is shown on the mean-field 
approximation of slave bosons that the rapid 
valence change can be realized for moderate strength of $\Ufc$ of the 
order of the bandwidth of conduction electrons if the level of f-electron 
$\epsilon_{\rm f}$ is tuned relative to the Fermi level.  
It is also shown by taking into account the fluctuations of slave bosons 
beyond the mean-field values that the superconductivity can be induced 
in the d-wave channel associated with the rapid valence change provided 
that the sherical condcution band is adopted.  

In \S2, we introduce the extended periodic Anderson model and briefly 
review the relevant works performed so far.  In \S3, we present 
a formalism for obtaining mean-field solutions in the slave-boson 
approach and taking into account the fluctuations around it.  
The results of mean-field solutions of slave bosons and 
superconducting transition temperature are presented in \S4.  
The main resutls of this paper are as follows: 
1) Sharp valence change is caused by the effect of $\Ufc$ with 
moderate strength of the order of the bandwith of conduction electrons, 
when the f-level $\epsilon_{\rm f}$ is tuned as a mimic of the effect of 
the pressure.  
2) The superconducting state is induced by 
the process of exchanging the slave-boson fluctuations for 
the values of $\epsilon_{\rf}$ at which the sharp valence change occurs.  
3) The symmetry of so induced superconducting state is $d$-wave if 
the spherical model is adopted for conduction electrons.  
The discussions are given in \S5.

\section{Extended Periodic Anderson Model with f-c Coulomb 
Interaction}
The effects of pressure appear through changes of the 
parameters characterizing the physics, the Kondo coupling $J$, 
the hybridization matrix element $V$, and the f-level $\ep_\rf$.  
However, these variations of parameters themselves cannot 
afford causing the drastic change of valence of Ce-ion 
from that of the Kondo regime to the VF 
regime, which leads to the divergent increase of the valnece 
susceptibility, while the valence increases prominently as the 
strength of $V$ is increased or the level $\ep_\rf$ is increased 
appraoching the Fermi level.~\cite{rf:haldane}  
Here we take into account the effect of the short range Coulomb 
repulsion between f electron and conduction electrons, because 
the repulsion may promote the charge transfer between both type of 
electrnic states in general when it is combined with the 
level shift of $\ep_\rf$.  Namely, we extend the conventional 
periodic Anderson model (PAM) as follows: 
\begin{eqnarray}
\sH
 = \sum_{k \sigma}(\ep_k-\mu) c_{k \sigma}^{\dagger}c_{k \sigma}^{}
 +\ep_\rf \sum_{k \sigma}f_{k \sigma}^{\dagger}f_{k \sigma}^{} 
 +U_{\rf\nolig\rf}\sum_i n_{i \uparrow}^\rf n_{i \downarrow}^\rf %
\nonumber \\
 +V\sum_{k \sigma}(c_{k \sigma}^{\dagger}f_{k \sigma}^{}+{\rm h.c.})
 +\Ufc\sum_{i \sigma \sigma'}n_{i \sigma}^{\rm f}n_{i \sigma'}^\rc,
 \label{eq:PAMUfc}
\end{eqnarray}
where the conventional notations for PAM are used except for $U_{\rm fc}$, 
the f-c Coulomb repulsion.  

The effect of $\Ufc$ on the valence of Ce-ion has been discussed in a 
number of contexts and models.  
First, with regard to the impurity Anderson model, the effect of 
$\Ufc$ was discussed in relation to optical experiments, 
valence-band photoemission (PES) and bremsstrahlung isochromat 
spectroscopy (BIS).~\cite{rf:Khomskii,rf:Alascio,rf:CH,rf:Hewson}
Costi and Hewson studied the effects of $\Ufc$ by a numerical 
renormalization group (NRG) approach.~\cite{rf:CH,rf:Hewson}
They concluded that in the Kondo regimes $\Ufc$ can be absorbed 
to other parameters ($V$, $\ep_\rf$ and $U_{\rf\nolig\rf}$), 
and the same set of renormalized parameters on 
$\Ufc=0$ is consistent with both the valence-band PES and BIS 
spectra and the thermodynamic properties. 
The effect of increasing $\Ufc$ is to increase the Kondo 
temperature, thus to decrease the number of f-electrons $n_\rf$ per 
ion mored rapidly.  
Indeed, a related model has been investigated on the basis of 
NRG approach by Takayama and Sakai, and it turned out that $n_\rf$ 
rapidly decreases at $\epsilon_{\rm f}\approx E_{\rm F}-\Ufc$, 
$E_{\rm F}$ being the Fermi level, as $\ep_\rf$ is increased 
so as to approach $E_{\rm F}$~\cite{rf:Takayama}. 
This result implies that the rapid valence change occurs where 
$\epsilon_{\rm f}+\Ufc$, the energy of $f^{1}$ state, and 
$E_{\rm F}$, the energy of the state $f^{0}$+(extra 
conduction electron on the Fermi level), are nearly degenerate 
giving rise to enhanced valence fluctuations.  
Then, $\Ufc$ has a considerable effect on the valence fluctuation, 
although it does not cause a valence instability in the 
impurity Anderson model.

On the other hand,  it is not easy to examine the effects of 
$\Ufc$ in the PAM due to the lattice effect.
In the case of impurity model, the conduction band plays a role 
of the electron bath so that the chemical potential $\mu$ is essentially 
fixed.  However, in the case of lattice model, 
$\mu$ is considerably affected by $\Ufc$ itself so that we have to 
treat the problem in a self-consistent fashion.
In particular, such a self-consistent treatment is indispensable 
when the valence state of Ce begins to leave from the Kondo regime 
to VF one under the high pressure.
There exist some studies of the extended PAM with $\Ufc$ within 
Hartree-Fock like 
approximations.~\cite{rf:Silva,rf:Hewson2,rf:Singh}
Even in this simple level of approximation, $\Ufc$ is responsible 
for the rapid change of the number of f-electrons as the level 
$\ep_\rf$ is tuned, leading to the first order transition in the 
large $\Ufc$ region.  
Recently, we have investigated the extended PAM with $\Ufc$ by the 
variational Monte Carlo method on the extended Gutzwiller variational 
wave function.~\cite{rf:YO1} 
It has been shown there that the valence of f-electrons decreases 
rapidly as the level of f-electrons $\epsilon_{\rm f}$ is increased, 
if $U_{\rm fc}$ is moderately large comparable to half of the bandwidth 
of conduction electrons.

The similar effects of the d-p Coulomb interaction in the so-called 
d-p model have been investigated as a possible charge fluctuation 
mechanism of the high $T_\rc$ 
superconductor.~\cite{rf:Valma-dp,rf:Hirashima}  
On the other hand, the physics discussed in the present paper 
is rather different from ``valence fluctuation mechanism" proposed 
for heavy electron superconductors in Ref.\ \citen{Brandow}.

On the basis of these considerations, we investigate the effect of 
$\Ufc$ in the extended PAM as a possible origin of the salient phenomena 
observed in \Ge under high pressures, particularly a possible 
mechanism of the enhancement of the superconducting 
transition temperature which has not been clarified so far.
We study the extended PAM with $\Ufc$ by 
the {\it slave-boson} and large-$N$ expansion 
approach~\cite{rf:RN1,rf:RN2,rf:ML}, which has been shown to be effective 
for studying thermodynamics properties, superconducting transition 
temperature, transport properties, etc. of the usual 
PAM~\cite{rf:RN1,rf:RN2,rf:ML,rf:AL,rf:LML,rf:HRW}.  
Following these methods, we extend it so as to include the f-c Coulomb 
interaction within a Gaussian fluctuation approximation and 
study its effects on the superconducting transition temperature.
\section{Formulation}
In this section we summarize a formulation for calculations 
using the {\it slave-boson} 
large-$N$ expansion technique~\cite{rf:RN1,rf:RN2,rf:ML}. 
We generalize the Hamiltonian eq.~\eqref{eq:PAMUfc} to the case 
where the  $N$-hold degeneracy exists , including both spin and 
orbital degrees of freedom. 
Then, we introduce the slave-boson operators to exclude 
the doublely occupied state by f-electrons assuming 
$U_{\rm f\nolig f}=\infty$.
Thus, the Hamiltonian is expressed as follows:
\begin{equation}
  \sH = \sum_{\bk m} \left( \ep_k c_{\bks m}^\dagger c_{\bks m}+ %
              \ep_{\rm f} f_{\bks m}^\dagger f_{\bks m}\right) + %
             \frac{V}{\sqrt{\NL}}\sum_{\bk\bq m}\left( %
              c_{\bks m}^\dagger f_{\bks+\bqs\,m} b_\bqs^\dagger + {\rm h. c.} \right)+%
             U_{\rm f\nolig c}\sum_{ilm}f_{il}^\dagger f_{il} %
                                      c_{im}^\dagger c_{im}, %
   \label{eq:Hsb}
\end{equation}
where $b$'s are slave-boson operators describing f$^{\,0}$ state and 
the constraint
\begin{equation}
  Q_i = \sum_{m} f_{im}^\dagger f_{im} + b_i^\dagger b_i = 1 %
   \label{eq:HsbQ}
\end{equation}
is required at each site in order to maintain an equivalence of 
the truncated 
Hilbert space and the original one.
To generate a $1/N$-expansion we rewrite variables as follows:
\begin{equation}
 \left\{
 \begin{array}{ccc}
  Q_i & \to & q_0 N \\
  b   & \to & b\sqrt{N} \\
  V   & \to & V/\sqrt{N} \\
  U_{\rm f\nolig c} & \to & U_{\rm f\nolig c}/\sqrt{N}
 \end{array}
 \right.
 \label{eq:Nscale}
\end{equation}
Hereafter, we use the radial gauge following Ref.\ \citen{rf:RN1}.  
(Although the radial and the Cartesian gauge formulation are ultimately 
equivalent, 
spurious infrared divergences do not appear in the radial 
gauge approach.~\cite{rf:ML}) 
We perform a local gauge transformation, 
$b_i(\tau)=\rho_i(\tau)\ex^{\mi \theta_i(\tau)}$, 
$f_{im}(\tau)=f'_{im}(\tau)\ex^{\mi \theta_i(\tau)}$ and 
$\lambda'_i(\tau)\equiv\lambda_i + \dot{\theta}_i(\tau)$. 
Then, the partition function is given 
by 
\begin{equation} \label{eq:Z1}
 Z=\int \sD (c c^\dagger f\nolig f^\dagger \rho \lambda) \exp(-S), 
\end{equation}
where
\begin{eqnarray}
  S&=&\int_0^\beta \rd\tau \Biggl[\;  %
   \sum_{\bk\bk'm} f_{\bks m}^\dagger(\tau) %
    \{(\partial_\tau + \ep_\rf)\delta_{\bks\bks'}+ \frac{1}{\sqrt{\NL}} %
    \mi \lambda(\bk-\bk';\tau) \} f_{\bks'm}(\tau)  \nonumber \\
   &+&\sum_{\bk m}c_{\bks m}^\dagger(\tau)(\partial_\tau+\ep_\bks)c_{\bks m}(\tau) 
    +\frac{V}{\sqrt{\NL}}\sum_{\bk\bk' m}\{c_{\bks m}(\tau)f_{\bks' m}(\tau)     \rho(\bk-\bk';\tau)+ {\rm h. c.} \}  \nonumber \\
   &+&\mi\frac{N}{\sqrt{\NL}}\sum_{\bk \bk'}\rho(-\bk';\tau)\lambda(\bk'-\bk;\tau) 
    \rho(\bk;\tau)-\mi q_0 N\sqrt{\NL}\lambda(\bzero;\tau) 
   +\frac{U_{\rm fc}}{N}\sum_{ilm}n_{il}^\rf n_{im}^\rc \Biggr].
 \label{eq:S1}
\end{eqnarray}
In the above expressions, we have rewritten variables $f'$ and $\lambda'$ 
as $f$ and $\lambda$, respectively, and neglected the Jacobian factor 
$\prod_{i\tau} \rho_i(\tau)$ following Ref.~\citen{rf:RN1}.

By introducing two kinds of Stratonovich-Hubbard fields $\phif$ and 
$\phic$ for $U_{\rm fc}$,
we can perform the functional integral over the Fermion fields in 
eq.~\eqref{eq:Z1} and eq.~\eqref{eq:S1}.  
Then, the partition function eq.~\eqref{eq:Z1} can be transformed 
as follows (See Appendix A~\ref{sec:appSHUfc} for a detail of derivation):  
\begin{equation}
 Z=\int \sD (\rho\lambda\phif\phic) \exp(-S), %
 \label{eq:Z2}
\end{equation}
\begin{eqnarray}
  S=&-&N\Tr\ln \hat{A} +\mi\frac{N}{\sqrt{\NL}}T^2\sum_{\rk\rk'}%
  \rho(-\rk)\lambda(\rk'-\rk)\rho(\rk) \nonumber \\%
   &-&\mi q_0 N\sqrt{\NL}\int \rd \tau \lambda(\bzero;\tau) %
    -\frac{NU_{\rf\rc}}{4}T\sum_\rk \phif(\rk)\phic(-\rk),
  \label{eq:S2}
\end{eqnarray}
where we have introduced abbreviation $\rk=(\bk,\mi \omega_n)$, etc. and 
matrix $\hat{A}$ is defined as
\begin{equation}
A_{\rk\rk'}= \left[
 \begin{array}{@{\,}c@{\,}}
  \,(-\mi\omega_n+\ep_\bks)\delta_{\rk\rk'}+%
     \frac{U_{\rf\rc}}{2\sqrt{\NL}}T\phif(\rk-\rk') \quad\qquad\qquad %
   T\frac{V}{\sqrt{\NL}}\rho(\rk-\rk') \qquad\\
  \, T\frac{V}{\sqrt{\NL}}\rho^\ast(\rk-\rk') \quad %
  (-\mi\omega_n+\ep_\rf)\delta_{\rk\rk'}+\frac{T}{\sqrt{\NL}}\mi\lambda(\rk-\rk') %
  +\frac{TU_{\rf\rc}}{2\sqrt{\NL}}\phic(\rk-\rk')
\end{array}
\right].
\end{equation}

At the level of mean-field approximation, the action $S_0$ can be written as
\begin{eqnarray}
 S_0&=&-N\Tr\ln \hat{A}_0 +\frac{\mi N}{T}%
      (\frac{\rhobar^2}{\sqrt{\NL}}-q_0\sqrt{\NL})\lambar%
       -\frac{NU_{\rf\rc}}{4}\frac{1}{T}\phif(\rk)\phic(-\rk) \label{eq:S0-1}\\
    &=&-N\sum_\rk\ln\left[ %
      (\mi\omega_n-\ep_\bks-\frac{U_{\rf\rc}}{2\sqrt{\NL}}\phifbar)%
      (\mi\omega_n-\ep_\rf-\frac{\mi\lambar}{\sqrt{\NL}}%
        -\frac{U_{\rf\rc}}{2\sqrt{\NL}}\phicbar)%
        -\frac{V^2}{\NL}\rhobar^2 \right] \nonumber \\
     &&\qquad %
      +N\frac{\mi}{T}(\frac{\rhobar^2}{\sqrt{\NL}}-q_0\sqrt{\NL})\lambar%
      -\frac{NU_{\rf\rc}}{4T}\phifbar\phicbar,
    \label{eq:S0-2}
\end{eqnarray}
where we have approximated $\lambda$, $\rho$ and $\varphi$ by their 
uniform mean field value as
\begin{equation}
 \left\{
  \begin{array}{@{\,}l}
   \lambda(\rmq)=\frac{1}{T}\lambar\delta_{\rmq} \\
   \rho(\rmq)=\frac{1}{T}\rhobar\delta_{\rmq} \\
   \varphi(\rmq)=\frac{1}{T}\bar{\varphi}\delta_{\rmq} \\
  \end{array}
 \right. ,
 \label{eq:valMF}
\end{equation}
and
\begin{eqnarray}
 A_{0\;\rk\rk'}&=&
 \left[
  \begin{array}{@{\,}c@{\,}}
  \,-\mi\omega_n+\ep_\bks+\frac{U_{\rf\rc}}{2\sqrt{\NL}}\phifbar \qquad %
   \frac{V}{\sqrt{\NL}}\rhobar \qquad\\
  \, \frac{V}{\sqrt{\NL}}\rhobar \quad %
  -\mi\omega_n+\ep_\rf+\frac{\mi\lambar}{\sqrt{\NL}} %
    +\frac{U_{\rf\rc}}{2\sqrt{\NL}}\phicbar 
  \end{array}
 \right]
 \delta_{\rk\rk'}\nonumber \\
 &\equiv& -\hat{G}_0^{-1}.
 \label{eq:A0-1}
\end{eqnarray}
Here $\hat{G}_0$ is nothing but a matrix of Green's functions for the 
renormalized band. 
In this mean-field approximation, quasiparticles acquire the 
{\it heavy effective mass} through the strong f-f Coulomb interactions, 
$U_{\rf\nolig\rf}=\infty$, and deep f-level $\ep_\rf$ far below the Fermi 
level.  From the optimum conditions for mean fields, 
$\partial S_0/\partial \rhobar=0$, 
$\partial S_0/\partial \lambar=0$ and 
$\partial S_0/\partial \bar{\varphi}=0$, 
we obtain the following set of self-consistent equations:
\begin{subeqnarray} \label{eq:mf}
  &&\frac{\mi \lambar}{\sqrt{\NL}} = -\frac{T}{\NL}\sum_\rk%
    \frac{V^2}{(\mi\omega_n-\epbar_\bk)(\mi\omega-\epbar_\rf)-\Vbar^2},%
    \label{eq:mf-a} \\
  &&q_0-(\frac{\rhobar}{\sqrt{\NL}})^2 = \frac{T}{\NL}\sum_\rk%
    \frac{\mi\omega_n-\epbar_\rk}%
    {(\mi\omega_n-\epbar_\bk)(\mi\omega-\epbar_\rf)-\Vbar^2},%
    \label{eq:mf-b} \\
  &&\frac{\phifbar}{2\sqrt{\NL}} = \frac{T}{\NL}\sum_\rk%
    \frac{\mi\omega_n-\epbar_k}%
    {(\mi\omega_n-\epbar_k)(\mi\omega-\epbar_\rf)-\Vbar^2},%
    \label{eq:mf-c} \\
  &&\frac{\phicbar}{2\sqrt{\NL}} = \frac{T}{\NL}\sum_\rk%
    \frac{\mi\omega_n-\epbar_\rf}%
    {(\mi\omega_n-\epbar_k)(\mi\omega-\epbar_\rf)-\Vbar^2},%
    \label{eq:mf-d}
\end{subeqnarray}
where we have introduced abbreviations as 
\begin{subeqnarray} \label{eq:bars}
 \epbar_\rk&\equiv&\ep_\rk+U_{\rf\rc}\nbar_\rf, \\ 
 \epbar_\rf&\equiv&\ep_\rf+\mi\lambar/\sqrt{\NL}+U_{\rf\rc}\nbar_\rc, \\
 \Vbar&\equiv& V\rhobar/\sqrt{\NL}, 
\end{subeqnarray}
where $\nbar_\rf\equiv\langle n_\rf \rangle_{\rm MF}$, the f-electron numbers 
per site and ``spin", and $\nbar_\rc\equiv\langle n_\rc \rangle_{\rm MF}$, 
the number of conduction electrons per site and ``spin".  
It is noted that both the left-hand sides of 
(\ref{eq:mf-b}b) and (\ref{eq:mf-c}c) are equal to $\nbar_\rf$ and 
that of (\ref{eq:mf-d}d) is equal to $\nbar_\rc$.  Thus, the following 
relations hold: 
\begin{eqnarray}
q_0-\left({\rhobar\over\sqrt{\NL}}\right)^2
&=&{\phifbar\over 2\sqrt{\NL}}=\nbar_\rf 
\label{eq:nof}\\
{\phicbar\over 2\sqrt{\NL}}&=&\nbar_\rc.
\label{eq:noc}
\end{eqnarray}
In terms of these quantities, the matrix Green function $\hat{G}_0$, 
eq.~\eqref{eq:A0-1}, can be expressed as 
\begin{eqnarray} 
 \hat{G}_0&=&\frac{1}{(\mi\omega_n-\epbar_\bks)(\mi\omega-\epbar_\rf)-\Vbar^2}
 \left[
  \begin{array}{@{\,}cc@{\,}}
  \mi\omega-\epbar_\rf & \Vbar \\
  \Vbar & \mi\omega-\epbar_\bks
  \end{array}
 \right]
\label{eq:G0mat}
\\
 &\equiv&
 \left[
  \begin{array}{@{\,}cc@{\,}}
  G_0^{\rc\rc} &  G_0^{\rc\rf} \\
  G_0^{\rf\rf} &  G_0^{\rf\rc} 
  \end{array}
 \right]
\label{eq:G0mat2}
\end{eqnarray}

\begin{figure}
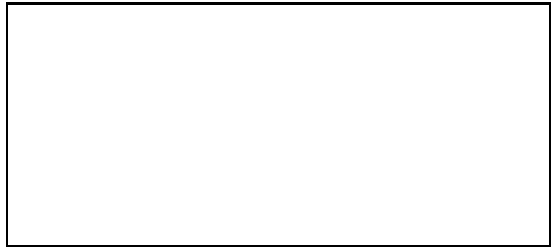

\figureheight{3cm}
  \caption{Feynman diagrams of interaction terms of eq.~\eqref{eq:Sf1}. %
  (a) represents the  hybridization process. (c), (d) and (g) are new terms %
  which appear through the f-c Coulomb interaction $\Ufc$.  (e) and (f) %
  represent the %
  second term of eq.~\eqref{eq:Sf1}, and (g) corresponds to the last term.}
  \label{fig:vertex}
\end{figure}

Next, we present the formalism treating the Gaussian fluctuations around 
their men-field values.  For this end, we express the boson fields as 
sums of the mean-field value (indicated by ``bar") 
and the fluctuations around it (indicated by "tilde") as
\begin{equation}
 \left\{
  \begin{array}{@{\,}l}
  \lambda(\rmq)=\frac{1}{T}\lambar\delta_{\rmq}+\lamtil(\rmq) \\
  \rho(\rmq)=\frac{1}{T}\rhobar\delta_{\rmq}+\rhotil(\rmq) \\
  \varphi(\rmq)=\frac{1}{T}\bar{\varphi}\delta_{\rmq}+\tilde{\varphi}(\rmq)
  \end{array}
 \right. .
 \label{eq:shift}
\end{equation}
Then, we expand $S$, (\ref{eq:S2}), with respect to the fluctuations 
up to the second order as 
\begin{equation}
S=S_0+\tilde{S},
\label{eq:S0plus1}
\end{equation} 
where $S_{0}$ is given by (\ref{eq:S0-2}) and 
the fluctuation part $\tilde{S}$ is given as follows:
\begin{equation}
 \tilde{S}=-N\Tr\ln\tilde{A}+\frac{\mi N}{\sqrt{\NL}}T\sum_\rk%
   \{ \lambar\rhotil(-\rk)\rhotil(\rk)+2\rhobar\lamtil(-\rk)\rhotil(\rk) \}
   -\frac{NU_{\rf\rc}}{4}T\sum_\rk\phiftil(\rk)\phictil(-\rk)
 \label{eq:Sf1}
\end{equation}
where 
$\tilde{A}\equiv1-\hat{G}_0 \hat{M}$ with 
$\hat{M}\equiv\hat{A}-\hat{A}_0$, the explicit form of which is 
\begin{equation}
M_{\rk\rk'}=
 \left[
  \begin{array}{@{\,}cc@{\,}}
  T\frac{U_{\rf\rc}}{2\sqrt{\NL}}\phiftil(\rk-\rk') & %
   T\frac{V}{\sqrt{\NL}}\rhotil(\rk-\rk') \\
  T\frac{V}{\sqrt{\NL}}\rhotil^\ast(\rk-\rk') &
   T\frac{\mi}{\sqrt{\NL}}\lamtil(\rk-\rk')+%
    T\frac{U_{\rf\rc}}{2\sqrt{\NL}}\phictil(\rk-\rk')
  \end{array}
 \right]. \label{eq:M}
\end{equation}
The first term of eq.~\eqref{eq:Sf1} can be expanded as 
\begin{equation}
-\Tr\ln\tilde{A}=-\Tr\ln(1-\hat{G}_0\hat{M})=%
 \sum_{n=1}^{\infty}\Tr(\hat{G}_0\hat{M})^n.
\end{equation}
We truncate this expansion at $n=2$ obtaining a Gaussian form.
Then, we can write the partition function of the Gaussian 
fluctuation part as (See Appendix~\ref{sec:appGF})
\begin{equation}
 \tilde{Z}=\int\sD(\rhotil\lamtil\phiftil\phictil)\exp[-\tilde{S}], \label{eq:Zf1}
\end{equation}
\begin{equation}
 \tilde{S}=NT\sum_\rk
(\rhotil(-\rk),\mi\lamtil(-\rk),\phictil(-\rk),\phiftil(-\rk))%
  \;\hat{S}_\rk%
  \;(\rhotil(\rk),\mi\lamtil(\rk),\phictil(\rk),\phiftil(\rk))^{t},%
 \label{eq:Sf2}
\end{equation}
where the superscript $t$ means the transpose is taken, and 
the symmetric matrix $\hat{S}_\rk$ is given as
\begin{eqnarray}
 \hat{S}_\rk&=&
 \left[
  \begin{array}{@{\,}cccc@{\,}}
  S_{\rr}(\rk) & S_{\rl}(\rk) & S_{\rpc}(\rk) & S_{\rpf}(\rk)     \\
  S_{\lr}(\rk) & S_{\lala}(\rk) & S_{\lpc}(\rk) & S_{\lpf}(\rk)   \\
  S_{\pcr}(\rk) & S_{\pcl}(\rk) & S_{\pcpc}(\rk) & S_{\pcpf}(\rk) \\
  S_{\pfr}(\rk) & S_{\pfl}(\rk) & S_{\pfpc}(\rk) & S_{\pfpf}(\rk)
  \end{array}
 \right], \label{eq:SfM1} \\
&=&
 \left[
  \begin{array}{@{\,}cccc@{\,}}
  \frac{\mi\lambar}{\sqrt{\NL}}+V^2\Picf(\rk)+V^2\Pith(\rk) &%
   \frac{\rhobar}{\sqrt{\NL}}+V\Pitw(\rk) &%
   \frac{U_{\rf\rc}}{2}V\Pitw(\rk) &%
   \frac{U_{\rf\rc}}{2}V\Pio(\rk) \\
  \star&\frac{1}{2}\Pif(\rk) &%
   \frac{\Ufc}{4}\Pif(\rk) &%
   \frac{\Ufc}{4}\Picf(\rk) \\
  \star&\star&\frac{\Ufc^2}{8}\Pif(\rk) &%
   -\frac{\Ufc}{8}+\frac{\Ufc^2}{8}\Picf(\rk) \\
  \star&\star&\star&\frac{\Ufc^2}{8}\Pic(\rk)
  \end{array}
 \right] ,\label{eq:SfM2}
\end{eqnarray}
where the definition of polarization functions $\Pi$'s are given in 
Appendix~\ref{sec:appGF}, and the symmetric off-giagonal components have 
abbriviated.  
The inverse matrix of $\hat{S}_\rk$ gives the matrix Green Functions for 
bose fields
\begin{equation}
D_{\alpha\beta}(\bk;\tau)\equiv%
  -\langle\mathrm{T}_\tau \alpha(\bk,\tau)\beta^\dagger(\bk,0)\rangle,
\label{eq:DMdef}
\end{equation}
where 
$\alpha,\;\beta$ are $\rhotil,\lamtil,\phiftil,\phictil$, as 
\begin{equation}
 \hat{D}(\rk)=-\hat{S}^{-1}(\rk). \label{eq:DM}
\end{equation}
In Fig.~\ref{fig:vertex}, we show the various possible interaction 
terms included in eq.~\eqref{eq:Sf1} 
by Feynman diagrams.
\section{Results}
\subsection{Mean-Field Solutions}
 First, we present the solutions within the mean-field approximation.  
Self-consistent equations \eqref{eq:mf} can be rearranged as 
follows~\cite{rf:Riseborough}:
\begin{subeqnarray} \label{eq:mf2}
 && \frac{\mi \lambar}{\sqrt{\NL}} =\frac{V^2}{\NL}\sum_\bk%
      \frac{f(E^-_\bks)-f(E^+_\bks)}{\sqrt{(\epbar_\rf-\epbar_\bks)^2+4\Vbar^2}} %
      \label{eq:mf2-a} \\
 &&q_0-(\frac{\rhobar}{\sqrt{\NL}})^2 =\frac{1}{2\NL}\sum_{\bk,\pm}%
      [1\pm\frac{\epbar_\rf-\epbar_\bks}%
           {\sqrt{(\epbar_\rf-\epbar_\bks)^2+4\Vbar^2}} ] f(E^\pm_\bks) %
      \label{eq:mf2-b} \\
 && \frac{1}{\NL}\sum_\bk\{f(E^-_\bks)+f(E^+_\bks)\}=\bar{n}_\rf+\bar{n}_\rc
\end{subeqnarray}
where $f(x)$ is the Fermi distribution function and $E^\pm$ are 
the quasi particle energies
\begin{equation}
E^\pm_{\bk}=\frac{1}{2}[\epbar_\rf+\epbar_\bks\pm\sqrt{%
                         (\epbar_\rf-\epbar_\bks)^2+4\Vbar^2 }].
\label{eq:Epm}
\end{equation}
A set of equations~\eqref{eq:mf2} is equivalent to that for the case of 
conventional PAM without $\Ufc$ except for the point that $\epbar_\bks$ and 
$\epbar_\rf$ include the term arising from $\Ufc$ as seen in ~\eqref{eq:bars}. 
We solve them for a three-dimensional model with the dispersion 
$\ep_\bks$ for the conduction-electrons approximated by that of 
free electron, $\ep_\bks=k^2/2m-D$, where the bottom of the 
conduction band is set as $-D$. 
Hereafter we use $D$ as the unit of energy. 
The bare mass, $m$, of conduction electrons is chosen 
such that the integration of, $\rho_0(\ep)$, 
the density of state per ``spin" of conduction electrons with respect to 
$\ep$ from $-D$ to $D$ is equal to 1: 
\begin{equation}
\rho_0(\ep)=\frac{3}{4\sqrt{2}D}\sqrt{\frac{\ep+D}{D}}.
\label{DOS0}
\end{equation}

\begin{figure}
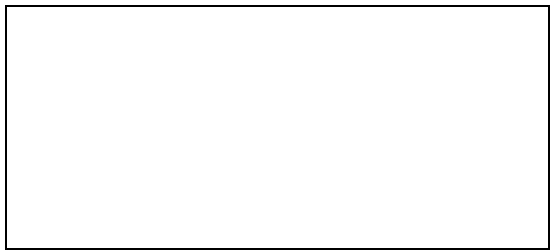

 \figureheight{3cm}
   \caption{$\bar{n}_\rf$, f-electron number per site and ``spin", %
   as a function of $\ep_\rf$, f-electron level, %
   in three-dimensinal model where the free electron %
    dispersion $\ep_\bks=k^2/2m-D$ is adopted for %
    the conduction electrons.}
   \label{fig:SBMF2}
\end{figure}

In Fig.~\ref{fig:SBMF2}, we show the results of 
$\bar{n}_\rf$ as a function of $\ep_\rf$ for several values of $\Ufc/D$.  
Here we set the hybridization as $V=$0.5$D$ and total electron number per 
``spin" as $n\equiv {\bar n}_{\rm f}+{\bar n}_{\rm c}=0.875$.  
These results are consistent with those of previous 
works.~\cite{rf:Silva,rf:Hewson2,rf:Singh}  
These results are consistent with a physical picture that the rapid 
valence change occurs at $\epsilon_{\rm f}\approx E_{\rm F}-\Ufc$ 
where the energies of $f^{0}$- and $f^{1}$-state 
are degenerate leading to enhanced valence fluctuations.
For much larger values of $\Ufc$ or smaller values of $V$ than those 
presented in Fig.~\ref{fig:SBMF2}, 
there occurs a first-order like discontinuous transitions although 
they are not shown in Fig.~\ref{fig:SBMF2}.
In the mean-field level of approximation for slave boson Hamiltonian, 
our treatment of the effect of $\Ufc$ is just like that in the 
Hartree-Fock approximation~\cite{rf:Silva,rf:Hewson2,rf:Singh}.
It is noted that the valence change occurs more sharply if we 
estimate it in much more proper approximation on the extended 
Gutzwiller variational wave function.~\cite{rf:YO1}  In this sense, the 
sharpness of the valence change may be underestimated by the present 
treatment.  

\begin{figure}
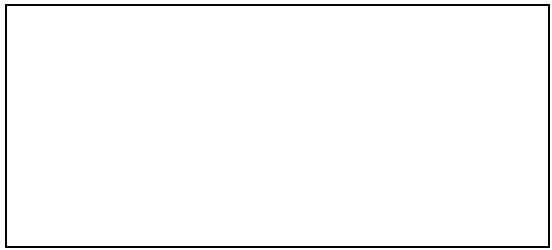

 \figureheight{3cm}
   \caption{Sommerfeld constant $\gamma$ vs Kondo temperature %
   $T_{\rm K}$, defined by $T_{\rm K}=\epbar_\rf-\mu$. %
   The parameters used are the same as those in Fig.~\ref{fig:SBMF2}. %
   Results for different values of $\Ufc/D$ lie on the same line, %
   exhibiting a universal behavior.}
   \label{fig:GvsTk}
\end{figure}

Using the mean-field solution we calculate the density of states 
$\rho(\mu)$ of quasiparticles at the Fermi energy.
It is related to the Sommerfeld constant $\gamma$ of the specific heat as 
\begin{equation} \label{eq:sommerfeld}
\gamma=\frac{\pi^2}{3}D\rho(\mu)N, 
\end{equation}
where $N$ is the degeneracy of the quasiparticles.  
The so-called Kondo temperature, or the characteristic temperature, 
of the present model, is defined simply as $(\epbar_\rf-\mu)$.  
In Fig.~\ref{fig:GvsTk}, we can see that the relation 
$T_{\rm K}\propto\gamma^{-1}$ holds.  
It is remarked that the relation $T_{\rm K}$ vs. $\gamma$ with 
different value of $\Ufc$ are lying on a line exhibiting a kind of 
universality.  It is suggested that the effect of $\Ufc$ can be 
absorbed into the other parameters of the conventional PAM 
at least in the mean-field level, like in the case of the single 
impurity Anderson model.
However, in the lattice case, the first-order like tansition occurs 
through the effects of $\Ufc$ even in the mean-field approximation.
This is considered to be a distinctive effect of $\Ufc$ in the PAM.

\subsection{Superconducting Transition Temperature}
\begin{figure}
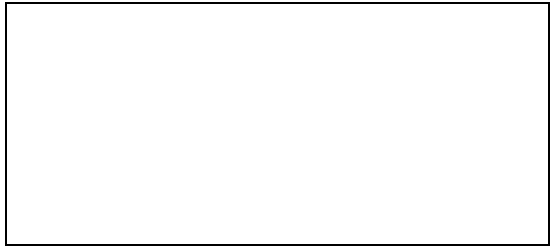

 \figureheight{3cm}
  \caption{Feynman diagram representing of the integral equations for %
  the scattering amplitude of two quasiparticles $\Gamma$ from %
  $(\mib{k},\mib{-k})$ to $(\mib{k}',\mib{-k}')$.}
  \label{fig:BCS}
\end{figure}

Next, we discuss a possible type of the superconducting gap near 
the region where the rapid valence change occurs owing to the f-c Coulomb 
interaction in the extended periodic Anderson model.
In the conventional PAM, problems of determining the superconducting 
transition temperature have been studied by several authors within the 
slave-boson and $1/N$-expansion method\cite{rf:LML,rf:HRW}.
Following their method, we calculate the superconducting transition 
temperature in the weak coupling theory in the present framework.
The transition temperature is obtained from the integral equation of 
the particle-particle scattering amplitude, $\Gamma$, for two 
quasiparticles with opposite momentum near the Fermi surface. 
(See Fig.~\ref{fig:BCS}.)
\begin{figure}
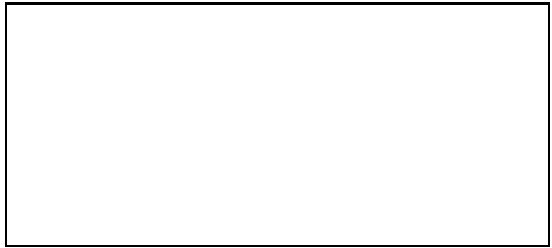

 \figureheight{3cm}
  \caption{Feynman diagram for the irreducible vertex part $\Gamma^{(0)}$ %
  to leading order in $1/N$. The explicit form of the fluctuation propagators %
  $D$ is given by eq.~\eqref{eq:DM} and its interaction vertices $g$ with %
  quasiparticle is shown in Fig.~\ref{fig:vertex}}
  \label{fig:gamma0}
\end{figure}
The quasiparticle operators can be represented by a unitary transformation 
in terms of f- and conduction electron operators as
\begin{equation}\label{eq:gammapm}
 \left[
  \begin{array}{@{\,}c@{\,}}
   \gamma^{(+)}_\bks \\ \gamma^{(-)}_\bks
  \end{array}
 \right]
=
 \left[
  \begin{array}{@{\,}cc@{\,}}
   u_\bks & v_\bks \\
   -v_\bks & u_\bks
  \end{array}
 \right]
 \left[
  \begin{array}{@{\,}c@{\,}}
   \rf_\bks \\ \rc_\bks
  \end{array}
 \right] ,
\end{equation}
where $\gamma^{(\pm)}$ corresponds to the eigen values $E^\pm$, 
eq.~\eqref{eq:Epm}, respectively.
The scattering amplitude $\Gamma$ is obtained from two-quasiparticle 
correlation function, 
\[ \langle{\rm T}_{\tau} \gamma^{(-)}_{\bks',m}(\tau_1) \gamma^{(-)}_{-\bks',m'}(\tau_2)%
\gamma^{(-)\dagger}_{-\bks,m'}(\tau_3) \gamma^{(-)\dagger}_{\bks,m}(\tau_4)\rangle, \] 
by removing the external legs.  

To leading order in $1/N$, irreducible vertex part $\Gamma^{(0)}$ 
includes only a single-boson exchange process as shown in 
Fig.~\ref{fig:gamma0}. 
With the use of the relations, eq.~\eqref{eq:gammapm}, eq.~\eqref{eq:DM}, 
and interaction vertices shown in Fig.~\ref{fig:vertex}, analytic expression 
of $\Gamma^{(0)}$ is given as follows:
\begin{eqnarray}
 \Gamma^{(0)}=&v^4&\left\{D_{\lala}+\frac{\Ufc}{2}D_{\lpc}+\frac{\Ufc}{2}D_{\pcl}+%
           (\frac{\Ufc}{2})^2D_{\pcpc}\right\} %
       -4uv^3\left\{VD_{\lr}+\frac{\Ufc}{2}D_{\pcr}\right\} \nonumber \\
       &+&u^2v^2\left[2\left\{\frac{\Ufc}{2}D_{\lpf}%
                      +(\frac{\Ufc}{2})^2D_{\pcpf}\right\}+4V^2D_{\rr}\right] \nonumber \\
       &-&u^3v4V\frac{\Ufc}{2}D_{\pfr} %
       +u^4(\frac{\Ufc}{2})^2D_{\pfpf} \label{eq:Gamma0}.
\end{eqnarray}
In the weak coupling limit, in which the external momenta are set on 
the Fermi surface, i.e. $|\bk|,|\bk'| \to k_{\rm F}$, and 
the static limit, $\omega \to 0$, is taken in the boson propagator, 
the linearized gap equation represented by Fig.~\ref{fig:BCS} 
can be solved by a conventional method.
\begin{figure}
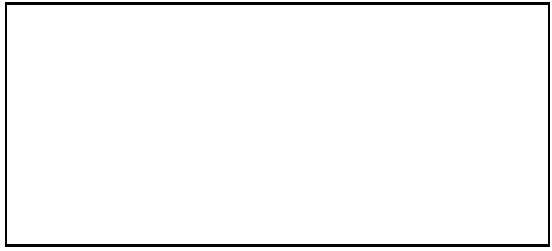

 \figureheight{3cm}
  \caption{$T_\rc$ for $d$-wave channel and ${\bar n}_\rf$, 
   f-electron number per site and ``spin", as a function of $\ep_\rf$.}
  \label{fig:Tcvsef}
\end{figure}
Namely, the scattering amplitude is decomposed into the Legendre 
polynomial as 
\begin{equation}
\Gamma(\bk,\bk^{\prime})
=\sum_{\ell=0}^{\infty}(2\ell+1)\Gamma_{\ell}
 \mathrm{P}_{\ell}(\hat{\bk}\cdot\hat{\bk}^{\prime}).
\end{equation} 
The scattering amplitude $\Gamma_{\ell}$ corresponding to the channel 
with relative angular momentum $\ell$ is given by
\begin{equation}
 \Gamma_{\ell}=\frac{\Gamma^{(0)}_{\ell}}{1+\rho(\mu)\Gamma^{(0)}_{\ell}%
   \ln(\frac{T_{\rm K}}{T})},
\end{equation}
where $\Gamma^{(0)}_{\ell}$'s are related to 
$\Gamma^{(0)}(\hat{\bk}\cdot\hat{\bk}^{\prime})$'s by the formula 
\begin{equation}
 \Gamma^{(0)}_{\ell}=\frac{1}{2}\int_{-1}^1 \rd(\cos\theta)%
   \Gamma^{(0)}(\theta){\rm P}_l(\cos\theta).
\end{equation}
Then the transition temperature $T_\rc$ for $\ell$-wave channel is given by
\begin{equation}
 T_\rc=T_{\rm K}
\exp\left[\frac{1}{\rho(\mu)\Gamma^{(0)}_{\ell}}\right].
\end{equation}
Here it is noted that the energy cut-off, corresponding to the 
Debye frequency, is given by $T_{\rm K}\equiv {\bar \epsilon}_{\rf}-\mu$, 
the bandwidth of quasiparticles.

\begin{figure}
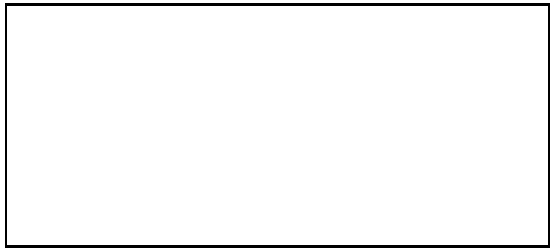

 \figureheight{3cm}
  \caption{$\Gamma^{(0)}(q)$ as a function of momentum transfer 
  $q$ for each scattering channel included in eq.~\eqref{eq:Gamma0}. 
  For example, ``ffff" means the $v^4$ term of eq.~\eqref{eq:Gamma0}. 
  The others are represented similarly.}
  \label{fig:fffc}
\end{figure}

The transition temperature $T_{\rm c}$ so calculated can exist 
only for the $d$-wave ($\ell=2$) channel 
as far as the channels, $\ell=0,1$ and $2$, are concerned. 
In Fig.~\ref{fig:Tcvsef} we show $T_{\rm c}$ as a function of 
$\epsilon_\rf$, the f-electron level relative to the Fermi level, 
for seeral values of $\Ufc$, the repulsion between f- and 
conduction electrons.  Parameters adopted are the same as those in 
Fig.\ 2.  There exists a sharp peak of $T_\rc$ at around $\ep_\rf$ 
where $n_\rf$ starts to show a rapid decrease.
Its tendency becomes more drastic as $\Ufc$ increases making 
the valence change sharper.
In the region where the f-electron number is decreased enough, 
$T_\rc$ is strongly suppressed.
We have investigated which term in eq.~\eqref{eq:Gamma0} plays 
important role for pairing interaction. 
It turned out that the major part of $\Gamma^{(0)}$ is induced 
by the scattering process (f,f)$\to$(f,c) or (f,c)$\to$(f,f), 
in which the valence of f-electrons is changed directly. 
(See fig.~\ref{fig:fffc}.)

In the Kondo limit where $\nbar_\rf\approx 1/2$, the spin fluctuations 
are believed to play the most important role for the occurrence of 
superconductivity.
In such a region we have to take into account the higher order term 
beyond $1/N$ to discuss the instability to the superconducting state, 
since the spin-fluctuation contribution to the effective interaction 
appears only beyond at the order $(1/N)^2$.~\cite{rf:HRW}
However, the present approach of the order of $1/N$ is still 
expected to work in the region where the valence fluctuations play 
an important role.
\section{Conclusions and Discussions}
We have developed a microscopic theory to support the idea that 
the rapid change of f-valence of Ce ion is the origin of anomalous 
behaviors of CeCu$_2$Ge$_2$ under the pressure $P\simeq 17$ GPa, 
such as the coincidence of peaks of $T_\rc$ and $\rho_0$, and 
the rapid decrease of the coefficient $A$ of $T^2$-term of the 
resistivity.  To this end, we analyzed the properties of an 
extended periodic Anderson model with the f-c Coulomb repulsion 
$\Ufc$ by the slave-boson and $1/N$-expansion method.  
It turned out that such a model indeed contains the ingredient 
making rapid change of $n_{\rm f}$, f-electron number per site, 
possible.  It was also shown that the superconducting transition 
temperature of $d$-wave pairing is sharply enhanced just before the rapid 
decrease of $n_{\rm f}$ as $\epsilon_{\rm f}$ is increased.  It turned out 
that the scattering process of (f,f)$\to$(f,c) or (f,c)$\to$(f,f), in 
which the f-electron number is changed directly, plays a major part 
in Cooper 
pair formation.  Cooper pairing of channels other than the d-wave 
cannot be possible so long as the pairing interaction discussed 
in the present paper is concerned.  

A few remarks should be made concerning the following points.  
First, although we did not mention the relation between the coefficient 
$A$ of $T^2$-term of the resistivity and the Sommerfeld constant 
$\gamma$ within the present approach, we have verified that 
the Kadowaki-Woods relation is reproduced in the heavy electron 
limit.~\cite{rf:YOD}
However, the crossover of the universal ratio 
$A/\gamma^{2}$ between classes of the heavy electron and of the 
transition metal was not reproduced.
We have cognizance of the necessity of calculations in which the 
dynamical effect of quasiparticle selfenergy is fully taken into account 
on the order of $1/N$.  However, we have to left such calculations 
for a future study because we need much more considerable amount of 
study for performing its program.  

Second, there exist several points of view different from ours for 
explaining the anomalous properties of \Ge.  
One of them is to attribute its origin to the orbital fluctuations 
in the multiband periodic Anderson model, in which the broad 
bandwidth under high pressure is expected to change 
the degeneracy of the f-electron state.  
Indeed this has been proposed in ref.~\citen{rf:Jac2} paying attention 
to the fact that, at the pressure corresponding to the peak of 
$T_\rc$ and $\rho_0$, the two temperatures $T_1^{\rm max}$ and 
$T_2^{\rm max}$ at which the resistivity takes peak merges with each other, 
indicating the Kondo temperature becomes 
of the order of the crystal field (CF) splitting.  
The orbital fluctuation mechanism has also been proposed as a possible 
mechanism for explaining the phenomena observed in CeCu$_2$Ge$_2$, 
and discussing that the orbital fluctuation is enhanced where 
$T_{\rm K}$ is comparable to CF splitting.~\cite{rf:Ikeda}
However, it is not clear whether all the anomalies observed in 
CeCu$_2$Ge$_2$ can be explained by those mechanisms.  

Third, the origin of superconductivity around the QCP is believed to be 
due to the enhanced spin fluctuations, while we have neglected here its
effect because we are interested in the physics in the region far from 
the QCP. 
Therefore, the primary origin of the superconducting transition 
may change to the spin-fluctuation mechanism as QCP is approached.
In order to discuss this crossover in the present model, we have to study 
the problem up to $(1/N)^2$ which is beyond scope of this paper.

Last, we have discussed the superconducting transition temperature 
in the weak coupling theory. However,  
it is suggested that the frequency dependence of the selfenergy 
is important as can be seen in the discussions concerning the 
Kadowaki-Woods relation~\cite{rf:YOD}.  
So, the study in the strong coupling theory of superconductivity 
is desired for more solid conclusion.  Nevertheless, the result does 
not seem to be modified qualitatively, judging from our past experience 
in the problem of Cooper pair formation in isotropic two-dimensional 
Fermions with short range repulsion~\cite{paramag2d}.

\section*{Acknowledgements}
One of the authors (Y.O.) acknowledges H. Maebashi and 
H. Kohno for their useful comments and valuable discussions.  
Y.O. was supported by Research Fellowships of the Japan Society for 
the Promotion of Science for Young Scientists.  
One of the authors (K.M.) acknowledges D. Jaccard for leading his 
attention to this interesting problem and O. Narikiyo for the 
stimulating discussions at early stage of this work.  K.M. also 
aknowledges a hospitality of J. Flouquet and his colleagues extended 
to him at CEA/Grenoble where the final form of the manuscripts 
was completed.  
This work is supported in part by a Grant-in-Aid for COE 
research (10CE2004) by Monbusho, the Ministry of 
Education, Science, Sports and Culture.
\appendix
\section{} \label{sec:appSHUfc}%
Here we discuss how the two kinds of boson fields are introduced in 
relation to the f-c Coulomb interaction.
The f-c Coulomb interaction term in eq.~\eqref{eq:Hsb} can be rewritten as 
\begin{equation}
 \sum_{i \alpha \beta} n_{i \alpha}^\rf n_{i \beta}^\rc= %
  \frac{1}{4}\sum_i [\{ \sum_\alpha (n_{i \alpha}^\rf + n_{i \alpha}^\rc)\}^2 - %
  \{ \sum_\alpha (n_{i \alpha}^\rf - n_{i \alpha}^\rc)\}^2 ].
\end{equation}
With the use of the identity 
\begin{eqnarray}
 \int \sD (\eta\,\zeta) \exp \Big[ \frac{\Ufc}{4N}\int_0^\beta \rd\tau 
 \{ (-\mi\eta_i(\tau)&+&\sum_\alpha (n_{i \alpha}^\rf + n_{i \alpha}^\rc) )^2 \nonumber \\
 &-&(-\zeta_i(\tau)+\sum_\alpha (n_{i \alpha}^\rf - n_{i \alpha}^\rc) )^2 \}%
   \Big] = {\rm Const}.
 \label{eq:appid1}
\end{eqnarray}
and eq.~\eqref{eq:appid1}, the term including $\Ufc$ in eq.~\eqref{eq:Z1} 
can be expressed as follows:
\begin{eqnarray}
  &\int \sD (\eta\,\zeta) \exp \Big[-\int_0^\beta \rd\tau \frac{\Ufc}{4N}%
   \sum_{\alpha \beta}n_{i \alpha}^\rf n_{i \beta}^\rc \Big] \nonumber \\
  &=\int \sD (c c^\dagger f f^\dagger \eta \zeta) \exp \Big[%
    -\int_0^\beta \rd\tau \{ \frac{\Ufc}{4N} (\eta_i^2(\tau)+\zeta_i^2(\tau)) \\
    &{}+\frac{\Ufc}{2N}((\mi\eta_i(\tau)-\zeta_i(\tau))\sum_\alpha n_{i\alpha}^\rf %
      +(\mi\eta_i(\tau)+\zeta_i(\tau))\sum_\alpha n_{i\alpha}^\rc) \} \Big]
  \\
   &=\int \sD (c c^\dagger f f^\dagger \phif \phic) \exp %
   \Big[-\int_0^\beta \rd\tau %
    \{ -\frac{N\Ufc}{4} (\phif_i(\tau)\phic_i(\tau) \nonumber \\
    &\quad\quad\quad+\frac{\Ufc}{2}(\phic_i(\tau)\sum_\alpha n_{i\alpha}^\rf +%
                    \phif_i(\tau)\sum_\alpha n_{i\alpha}^\rc) )\} \Big]. 
\end{eqnarray}
In the last line, bose fields, $\eta$ and $\zeta$, have been 
transformed to the new ones,
$\phif_i=N(\mi\eta_i+\zeta_i)$ and $\phic_i=N(\mi\eta_i-\zeta_i)$.
After carrying out the functional integral over the Fermion fields, 
we obtain eqs.~\eqref{eq:Z2} and \eqref{eq:S2}.
\section{} \label{sec:appGF}%
Here we derive formal expressions of the Gussian fluctuation part of 
the action, eq.~\eqref{eq:Sf1}, and 
symmetry property of the polarization functions, eq.~\eqref{eq:SfM2}.
Of the terms in the part of action, eq.~\eqref{eq:Sf1}, 
\[-\Tr\ln\tilde{A}=-\Tr\ln(1-\hat{G}_0\hat{M})=%
 \sum_n \Tr(\hat{G}_0\hat{M})^n, \] 
those with $n=2$ give contributions of the Gaussian fluctuation.
Namely, the relevant term is expressed as
\begin{equation}
 \frac{1}{2}\Tr(\hat{G}_0\hat{M})^2=\sum_{\rk\rmq}{\mathrm{tr}}\{%
 \hat{G}_0(\rk)\hat{M}_{\rk\;\rk+\rmq}\hat{G}_0(\rk+\rmq)\hat{M}_{\rk+\rmq\;\rk} \}.
 \label{eq:ln-expand}
\end{equation}
The explicit form of $\hat{G}_0\hat{M}$ is given as
\begin{eqnarray}
 &&\hat{G}_0 \hat{M}(\rk,\rk+\rmq)= \\
 &&\frac{T}{\sqNL}
 \left[
 \begin{array}{@{\,}cc@{\,}}
  \frac{\Ufc}{2}\Gcc(\rk)\phiftil(-\rmq)+V\Gcf(\rk)\rhotil(-\rmq) &%
   V\Gcc(\rk)\rhotil(\rmq)+%
   \Gcf(\rk)(\mi\lamtil(-\rmq)+\frac{\Ufc}{2}\phictil(-\rmq)) \\
  \frac{\Ufc}{2}\Gfc(\rk)\phiftil(-\rmq)+V\Gff(\rk)\rhotil(-\rmq) &
   V\Gfc(\rk)\rhotil(-\rmq)+%
   \Gff(\rk)(\mi\lamtil(-\rmq)+\frac{\Ufc}{2}\phictil(-\rmq)) 
 \end{array}
 \right] . \nonumber
\end{eqnarray}
Then, the diagonal components of the product of four matrices in 
eq.~\eqref{eq:ln-expand} are written as follows:
\begin{eqnarray}
  && [\hat{G}_0 \hat{M}(\rk,\rk+\rmq)\hat{G}_0 \hat{M}(\rk+\rmq,\rk)]_{(1\,1)}= \\
   &&\frac{T^2}{\NL}\Bigl(%
     \{ \frac{\Ufc}{2}\Gcc(\rk)\phiftil(-\rmq)+V\Gcf(\rk)\rhotil(-\rmq) \}%
     \{ \frac{\Ufc}{2}\Gcc(\rk+\rmq)\phiftil(\rmq)+V\Gcf(\rk+\rmq)\rhotil(\rmq) \} \nonumber \\
    &&+\{ V\Gcc(\rk)\rhotil(-\rmq)+%
        \Gcf(\rk)(\mi\lamtil(-\rmq)+\frac{\Ufc}{2}\phictil(-\rmq)) \}%
     \{ \frac{\Ufc}{2}\Gfc(\rk+\rmq)\phiftil(\rmq)+V\Gff(\rk+\rmq)\rhotil(\rmq) \}%
   \Bigr), \nonumber
 \end{eqnarray}
and
 \begin{eqnarray}
 && [\hat{G}_0 \hat{M}(\rk,\rk+\rmq)\hat{G}_0 \hat{M}(\rk+\rmq,\rk)]_{(2\,2)}= \\
   &&\frac{T^2}{\NL}\Bigl(%
  \{ \frac{\Ufc}{2}\Gfc(\rk)\phiftil(-\rmq)+V\Gff(\rk)\rhotil(-\rmq) \} \nonumber \\%
   &&  \{ V\Gcc(\rk+\rmq)\rhotil(\rmq)+%
        \Gcf(\rk+\rmq)(\mi\lamtil(\rmq)+\frac{\Ufc}{2}\phictil(\rmq)) \} \nonumber \\
   && +\{ V\Gfc(\rk)\rhotil(-\rmq)+%
        \Gff(\rk)(\mi\lamtil(-\rmq)+\frac{\Ufc}{2}\phictil(-\rmq)) \} \nonumber \\
   &&  \{ V\Gcc(\rk+\rmq)\rhotil(\rmq)+%
        \Gcf(\rk+\rmq)(\mi\lamtil(\rmq)+\frac{\Ufc}{2}\phictil(\rmq)) \} %
   \Bigr). \nonumber
 \end{eqnarray}
Substituting these expressions into eq.~\eqref{eq:ln-expand} and 
making rather tedious rearrangements, we obtain eq.~\eqref{eq:Sf2} 
with the matrix $\hat{S}$ whose components are defined by 
eqs.~\eqref{eq:SfM1} and~\eqref{eq:SfM2}.
Polarization function $\Pi$'s in eq.~\eqref{eq:SfM2} are defined as follows:
\begin{equation}
 \left\{ %
 \begin{array}{ccl}
  \Pic(k,\mi\omega)&\equiv&\Pi^{(\rc\rc,\rc\rc)}(\bk,\mi\omega) \\
  \Pif(k,\mi\omega)&\equiv&\Pi^{(\rf\rf,\rf\rf)}(\bk,\mi\omega) \\
  \Picf(k,\mi\omega)&\equiv&\Pi^{(\rc\rf,\rc\rf)}(\bk,\mi\omega) \\
  &&\\
  \Pio(k,\mi\omega)&\equiv& \frac{1}{2}\{\Pi^{(\rc\rc,\rc\rf)}(\bk,\mi\omega)+%
                             \Pi^{(\rc\rf,\rc\rc)}(\bk,\mi\omega) \}\\
  \Pitw(k,\mi\omega)&\equiv& \frac{1}{2}\{\Pi^{(\rf\rf,\rc\rf)}(\bk,\mi\omega)+%
                             \Pi^{(\rc\rf,\rf\rf)}(\bk,\mi\omega)\} \\
  \Pith(k,\mi\omega)&\equiv& \frac{1}{2}\{\Pi^{(\rc\rc,\rf\rf)}(\bk,\mi\omega)+%
                             \Pi^{(\rf\rf,\rc\rc)}(\bk,\mi\omega)\}
 \end{array}
 \right. 
\end{equation}
where $\Pi^{(\alpha\beta,\gamma\delta)}$ is defined as 
\begin{equation}
 \Pi^{(\alpha\beta,\gamma\delta)}(k,\mi\omega)\equiv%
  \frac{T}{\NL}\sum_{\bq,\mi\nu}G_0^{\alpha\beta}(\bq,\mi\nu)%
          G_0^{\gamma\delta}(\bq+\bk,\mi\nu+\mi\omega), %
\end{equation}
where $\alpha,\beta,\gamma$ and $\delta$ stand for c or f.  
It is noted that $\Pi^{(\alpha\beta,\gamma\delta)}$ satisfies the 
following symmetry relations:
\begin{equation}
 \left\{ %
 \begin{array}{ccl}
 \Pi^{(\rc\rc,\rc\rf)}(\bk,\mi\omega) &=& 
                             \Pi^{(\rc\rf,\rc\rc)}(\bk,-\mi\omega) \\
 \Pi^{(\rf\rf,\rc\rf)}(\bk,\mi\omega)&=& 
                            \Pi^{(\rc\rf,\rf\rf)}(\bk,-\mi\omega) \\
 \Pi^{(\rc\rc,\rf\rf)}(\bk,\mi\omega)&=& 
                             \Pi^{(\rf\rf,\rc\rc)}(\bk,-\mi\omega).
 \end{array}
 \right.
\end{equation}

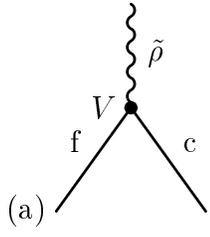
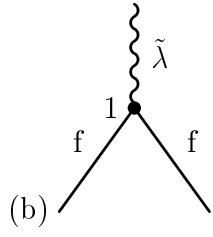
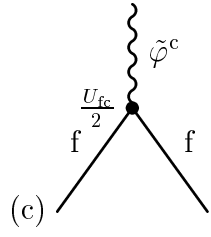
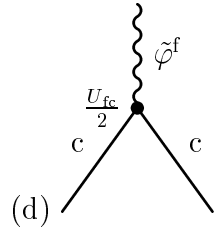
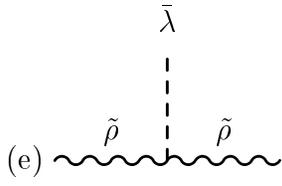
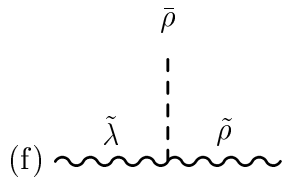
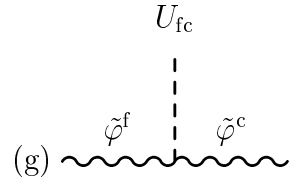
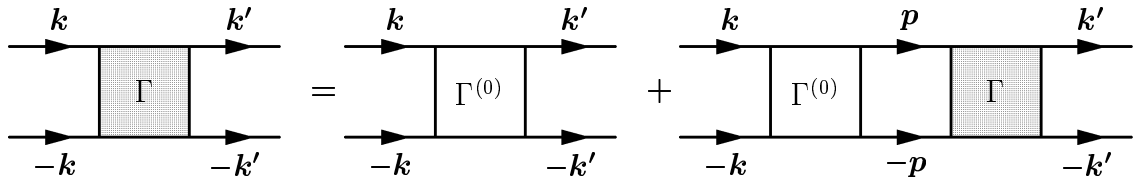
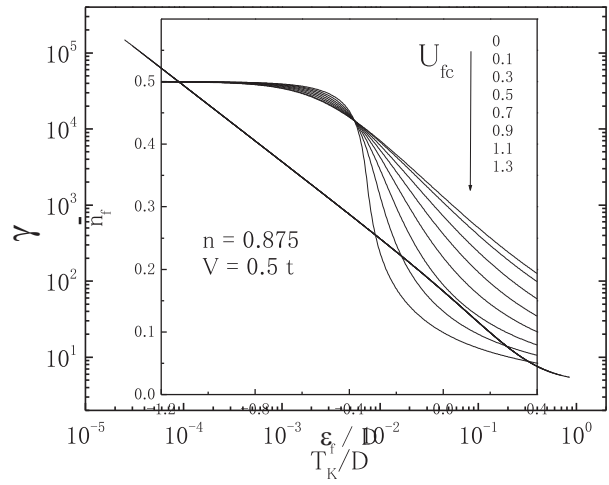
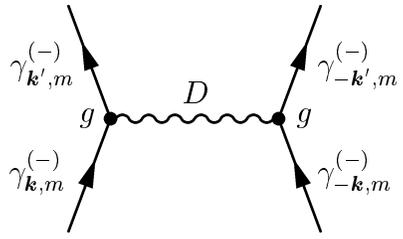
\end{document}